\documentclass[aps,prb,twocolumn,superscriptaddress]{revtex4-1}
\pdfoutput=1

\newcommand{\figurescale}{1}

\usepackage{verbatim}
\usepackage{amsmath}
\usepackage[utf8]{inputenc}
\usepackage[pdftex]{graphicx}
\usepackage[separate-uncertainty=true]{siunitx}
\DeclareSIUnit{\rpm}{rpm}
\usepackage[colorlinks=true,urlcolor=blue,linkcolor=blue,citecolor=blue]{hyperref}
\usepackage{xcolor}
\usepackage[T1]{fontenc}
\usepackage{placeins}
\usepackage{nicefrac}
\usepackage{braket}
\usepackage{pdfcomment}
\usepackage{xcolor}

\usepackage{lineno}
\begin{document}

%############################## TITLE #########################################
\title{Robust Valley Polarization of Helium Ion Modified Atomically Thin MoS$_2$}
%##############################################################################
%
%############################ AUTHORS #########################################
\author{J.~Klein}\email{julian.klein@wsi.tum.de}
\affiliation{Walter Schottky Institut and Physik Department, Technische Universit\"at M\"unchen, Am Coulombwall 4, 85748 Garching, Germany}
\affiliation{Nanosystems Initiative Munich (NIM), Schellingstr. 4, 80799 Munich, Germany}
\author{A.~Kuc}
\affiliation{Wilhelm-Ostwald-Institut f\"ur Physikalische und Theoretische Chemie, University of Leipzig, Linn\'estr.2, D-04103 Leipzig, Germany}
\affiliation{Department of Physics \& Earth Sciences, Jacobs University Bremen, Campus Ring 1, 28759 Bremen, Germany}
\author{A.~Nolinder}
\affiliation{Walter Schottky Institut and Physik Department, Technische Universit\"at M\"unchen, Am Coulombwall 4, 85748 Garching, Germany}
\author{M.~Altzschner}
\affiliation{Walter Schottky Institut and Physik Department, Technische Universit\"at M\"unchen, Am Coulombwall 4, 85748 Garching, Germany}
\author{J.~Wierzbowski}
\affiliation{Walter Schottky Institut and Physik Department, Technische Universit\"at M\"unchen, Am Coulombwall 4, 85748 Garching, Germany}
\author{F.~Sigger}
\affiliation{Walter Schottky Institut and Physik Department, Technische Universit\"at M\"unchen, Am Coulombwall 4, 85748 Garching, Germany}
\author{F.~Kreupl}
\affiliation{Department of Hybrid Electronic Systems, Technische Universit\"at M\"unchen, Arcisstr. 21, 80333 Munich, Germany}
\author{J.~J.~Finley}
\affiliation{Walter Schottky Institut and Physik Department, Technische Universit\"at M\"unchen, Am Coulombwall 4, 85748 Garching, Germany}
\affiliation{Nanosystems Initiative Munich (NIM), Schellingstr. 4, 80799 Munich, Germany}
\author{U.~Wurstbauer}
\affiliation{Walter Schottky Institut and Physik Department, Technische Universit\"at M\"unchen, Am Coulombwall 4, 85748 Garching, Germany}
\affiliation{Nanosystems Initiative Munich (NIM), Schellingstr. 4, 80799 Munich, Germany}
\author{A.~W.~Holleitner}\email{holleitner@wsi.tum.de}
\affiliation{Walter Schottky Institut and Physik Department, Technische Universit\"at M\"unchen, Am Coulombwall 4, 85748 Garching, Germany}
\affiliation{Nanosystems Initiative Munich (NIM), Schellingstr. 4, 80799 Munich, Germany}
\author{M.~Kaniber}\email{michael.kaniber@wsi.tum.de}
\affiliation{Walter Schottky Institut and Physik Department, Technische Universit\"at M\"unchen, Am Coulombwall 4, 85748 Garching, Germany}
\affiliation{Nanosystems Initiative Munich (NIM), Schellingstr. 4, 80799 Munich, Germany}
%
%##############################################################################

%
%##############################################################################
%									ABSTRACT
%##############################################################################
%
\begin{abstract}
\textbf{
Atomically thin semiconductors have dimensions that are commensurate with critical feature sizes of future optoelectronic devices defined using electron/ion beam lithography. Robustness of their emergent optical and valleytronic properties is essential for typical exposure doses used during fabrication. Here, we explore how focused helium ion bombardement affects the intrinsic vibrational, luminescence and valleytronic properties of atomically thin MoS$_{2}$. By probing the disorder dependent vibrational response we deduce the interdefect distance by applying a phonon confinement model. We show that the increasing interdefect distance correlates with disorder-related luminscence arising $\SI{180}{\milli\electronvolt}$ below the neutral exciton emission. We perform ab-initio density functional theory of a variety of defect related morphologies, which yield first indications on the origin of the observed additional luminescence. Remarkably, no significant reduction of free exciton valley polarization is observed until the interdefect distance approaches a few nanometers, namely the size of the free exciton Bohr radius. Our findings pave the way for direct writing of sub-10 nm nanoscale valleytronic devices and circuits using focused helium ions.
}
\end{abstract}
%
%##############################################################################
\maketitle
%
%###############################################################################
%								MAIN TEXT
%###############################################################################
%

Monolayer thick transition metal dichalcogenides (TMDCs) like MoS$_{2}$ or WSe$_{2}$ offer excellent perspectives for future application in a variety of optical and optoelectronic devices owing to their strong light-matter interaction and direct bandgap. \cite{Mak.2010,Splendiani.2010} Weak dielectric screening in the single layer limit promotes strong excitonic effects including large exciton binding energies with an energy structure that follows a non-hydrogenic Rydberg series. \cite{Chernikov.2014} The naturally broken inversion symmetry of monolayer MoS$_{2}$ provides optical access to address the valley and spin degrees of freedom using helical optical excitation \cite{Mak.2012,Zeng.2012,Cao.2012}; one of the basic requirements for future valleytronic applications.

The large surface-to-volume ratio of two-dimensional (2D) crystals offer various routes for modifying their unique photophysical \cite{Tongay.2013,Tongay.2013b,Nan.2014,Chow.2015} and electronic \cite{GhorbaniAsl.2013,Hong.2015,Li.2016,Ye.2016} properties to provide novel functionalities. In particular, it has been shown that physisorbed \cite{Tongay.2013,Miller.2015} and chemisorbed \cite{Nan.2014,Chow.2015} atoms and molecules onto pristine and defective MoS$_{2}$ and other TMDCs can modify their optical luminescence properties. For monolayer MoS$_{2}$, additional luminescence due to physisorbed N$_{2}$ molecules at disulphur vacancies V$_{2S}$ has been demonstrated to arise from $\alpha$ particle irradiation. \cite{Tongay.2013} Moreover, extrinsic modifications of the host crystal by annealing, \cite{Nan.2014} ion bombardement, \cite{Tongay.2013,Mignuzzi.2015,Fox.2015,Iberi.2016} electron beam irradiation \cite{Komsa.2012,Parkin.2016} or plasma treatment \cite{Nan.2014} have been reported. In particular, structuring and lithography with a focused He$^{+}$ beam is of particular interest owing to sub-nm beam diameters giving the opportunity for manipulation of 2D materials on the nanometer scale. In this regard tailoring of optically active defect centers in TMDCs, to even host single photon emitters as recently observed in hBN is highly desirable. \cite{Tran.2015}

%
%###################### Figure 1 ################################################
\begin{figure*}[!ht]
\scalebox{\figurescale}{\includegraphics[width=1\linewidth]{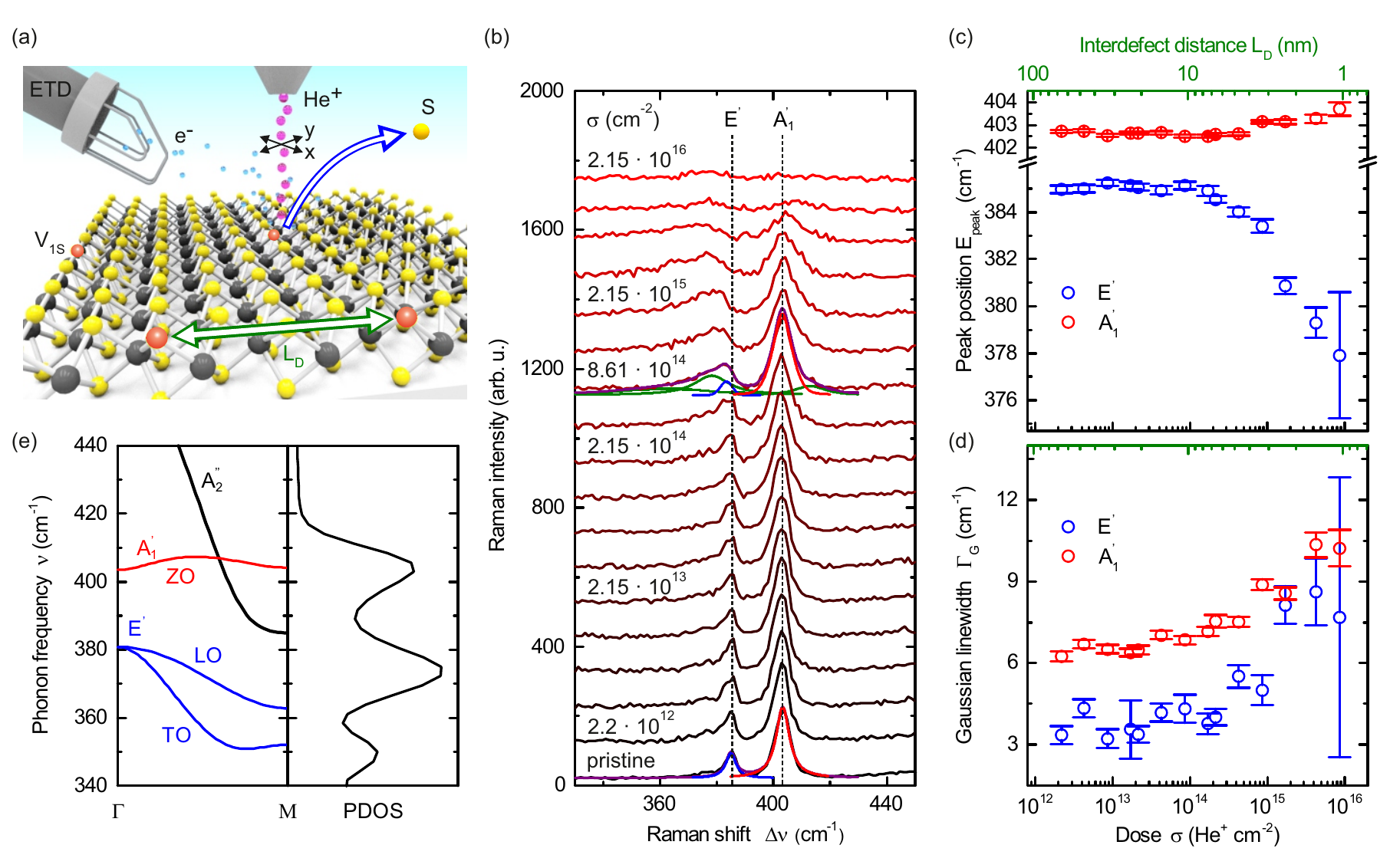}}
\renewcommand{\figurename}{Fig.}
\caption{\label{fig1}
\textbf{Phonon confinement in HIM structured MoS$_{2}$.}
(a) Schematic illustration of He$^{+}$ induced $V_{1S}$ in atomically thin MoS$_{2}$. The green arrow illustrates the interdefect distance $L_{D}$.
(b) Raman spectra of 1L MoS$_{2}$ at room temperature for increasing He$^{+}$ dose. The position of the out-of-plane $A^{'}_{1}$ and in-plane $E^{'}$ first-order phonon modes for pristine MoS$_{2}$ are marked by dashed lines. The data are displaced for clarity.
(c) Dose dependent evolution of peak position $E_{peak}$ of $A^{'}_{1}$ and $E^{'}$ deduced from fitting the data in Fig.~\ref{fig1}b (exemplarily shown as green lines for a dose of $8.61 \cdot 10^{14}$ $\SI{}{\per\centi\meter\squared}$)
(d) Dose dependent change in Gaussian linewidth $\Gamma_{G}$ of $A^{'}_{1}$ and $E^{'}$.
(e) Calculated phonon dispersion and corresponding phonon density of states (PDOS) of pristine 1L MoS$_{2}$ between $\Gamma$ and M point featuring LO, TO and ZO modes of $E^{'}$ (blue) and $A^{'}_{1}$ (red).
}
\end{figure*}

Here, we control the amount of disorder by utilizing a scanning helium ion microscope (HIM) to introduce defects within the monolayer crystal (Fig.~\ref{fig1}a). We probe the Raman signal of the first-order modes to quantify the HIM induced disorder and deduce an interdefect distance from distinct mode shifts using a phonon confinement model. \cite{Mignuzzi.2015,Shi.2016} We observe a significant disorder-related luminescence, which is redshifted by $\Delta E \sim \SI{180}{\milli\electronvolt}$ ($\Delta E \sim \SI{150}{\milli\electronvolt}$) with respect to the neutral exciton X (trion T) emission. In contrast to previous studies,\cite{Tongay.2013} the disorder-related luminescence is interpreted to originate from chemisorbed atoms/molecules at sulphur vacancies as bombardment-induced luminescence is observed at low-temperature ($\SI{10}{\kelvin}$) and high-vacuum ($10^{-6}\SI{}{\milli\bar}$) with mostly dose independent free exciton emission. Our findings are supported by ab-initio DFT calculations considering a variety of possible substituents, based on atoms and molecules at monosulphur vacancies. The disorder-related luminescence peak shows no valley polarization for quasi-resonant excitation of the neutral exciton. In strong contrast, we observe very robust valley polarization of the neutral exciton X and trion T of MoS$_{2}$, which is not systematically affected by HIM structuring up to doses of $\sim 5 \cdot 10^{13}$ $\SI{}{\per\centi\meter\squared}$. This finding further supports our working hypothesis that disorder-related emission predominantly stems from monosulphur vacancies $V_{1S}$ due to their specific symmetry resulting in weak intervalley scattering. \cite{Kaasbjerk.2016} \\

%
%###################### Figure 2 ################################################
\begin{figure*}[ht]
\scalebox{\figurescale}{\includegraphics[width=1\linewidth]{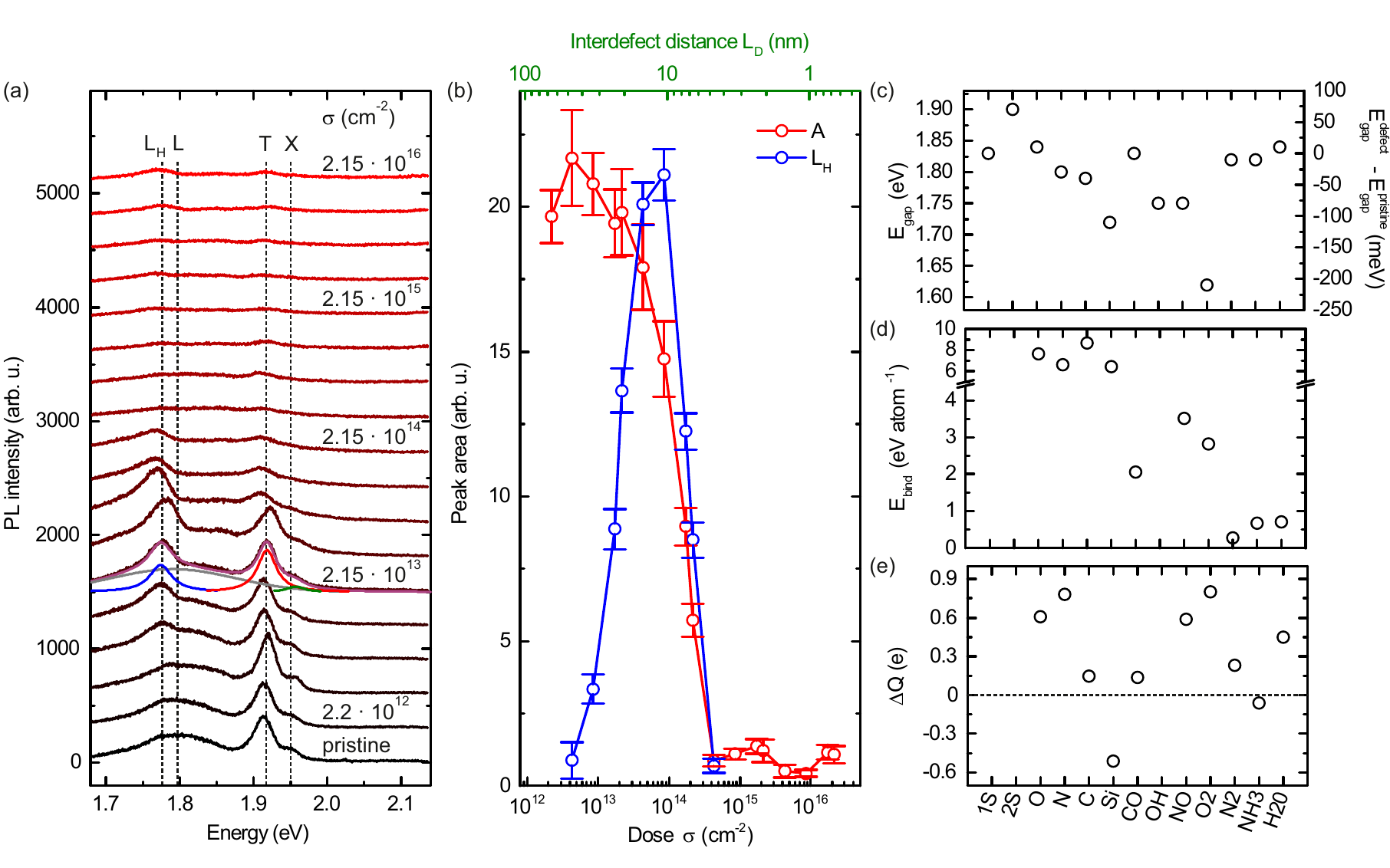}}
\renewcommand{\figurename}{Fig.}
\caption{\label{fig2}
\textbf{Luminescence modification of HIM structured MoS$_{2}$.}
(a) Low-temperature ($\SI{10}{\kelvin}$) $\mu$-PL spectra of 1L MoS$_{2}$ as a function of He$^+$ dose. Neutral exciton $X$, trion $T$ and low energy disorder-related peaks $L$ and $L_{H}$ are marked by dashed lines. A typical fit for a dose of $2.15 \cdot 10^{13} \SI{}{\per\centi\meter\squared}$ reveals the different contributions of $X$ (green), $T$ (red), $L$ (grey) and $L_{D}$ (blue). All spectra are vertically displaced for clarity.
(b) Fitted peak areas of $A = X + T$ and $L_{H}$ as a function of $\sigma$.
(c) Left scale: DFT calculated optical gap energies for the corresponding vacancies, adsorbed atoms and molecules. Right scale: Relative change in optical gap energy $\Delta E = E_{gap}^{defect} - E_{gap}^{pristine}$.
(d) DFT calculation of corresponding binding energies and (e) effective electron charge transfer of different atoms and molecules that are chemisorbed at monosulphur vacancies $V_{1S}$. The DFT calculation is performed for a $5\times5$ supercell with a single $V_{1S}$ corresponding to a vacancy density of $\sim 7 \cdot 10^{13} \SI{}{\per\centi\meter\squared}$.
}
\end{figure*}
%##############################################################################

\section{Results}

\textbf{Disorder induced modification of the MoS$_{2}$ phonon modes.} The monolayer MoS$_{2}$ crystal studied in this letter is mechanically exfoliated onto p$^{++}$ Si substrates covered with a \SI{285}{\nano\metre} thick layer thermally grown SiO$_{2}$. For introducing defects (disorder) we use a HIM, which enables irradiation of the TMDC with He$^{+}$ and a small beam diameter of < $\SI{1}{\nano\meter}$ (Fig.~\ref{fig1}a and Method Section). \cite{Fox.2015} Subsequent to He$^{+}$ exposure, we probe the dose dependent evolution of the first-order $A^{'}_{1}$ and $E^{'}$ Raman modes (Fig.~\ref{fig1}b). We analyze each Raman spectrum by fitting using pseudo-Voigt functions to each peak (see Method section). Typical fits for pristine MoS$_{2}$ and a dose of $8.61 \cdot 10^{14} \SI{}{\per\centi\meter\squared}$ are shown in Fig.~\ref{fig1}b. The homogeneous line broadening $\Gamma_{L}$ for the $A^{'}_{1}$ and $E^{'}$ Raman modes are used as fixed parameters that are obtained from fitting the pristine spectrum. Fig.~\ref{fig1}b shows the dose dependent trend of the first-order Raman peak positions. From our data, we observe two dissimilar dependences. For doses $10^{12}$ $\SI{}{\per\centi\meter\squared}$ < $\sigma$ < $10^{14}$ $\SI{}{\per\centi\meter\squared}$, no significant changes in the peak positions are observed (Fig.~\ref{fig1}c), while for doses $\sigma$ > $10^{14}$ $\SI{}{\per\centi\meter\squared}$ a maximum shift of $\SI{-7}{\per\centi\meter}$ ($\SI{1.2}{\per\centi\meter}$) of $E^{'}$ ($A^{'}_{1}$) is observed with a trend consistent with previous work where Mn$^{+}$\cite{Mignuzzi.2015} and Ar$^{+}$\cite{Shi.2016} are used for ion bombardement of MoS$_{2}$. Based on our results for $\sigma$ < $10^{14}$ $\SI{}{\per\centi\meter\squared}$, higher than typical doses used for HIM lithography,\cite{Sidorkin.2009} we can exclude strain and/or doping effects since the Raman peak positions are constant within the fitting error. Using the maximum fit error $\sim \SI{0.2}{\per\centi\meter}$ of the $E^{'}$ and $\sim \SI{0.08}{\per\centi\meter}$ for the $A^{'}_{1}$  mode, we can set an upper limit on strain $< \SI{0.2}{\percent}$ \cite{Rice.2013} and change in doping density to be $\Delta n < 10^{11} \SI{}{\per\centi\meter\squared}$.\cite{Chakraborty.2012} Moreover, for an increasing dose, the spectral linewidth is accompanied by an increase of the inhomogeneous (Gaussian) linebroadening $\Gamma_{G}$ by $\sim \SI{5}{\per\centi\meter}$ ($\sim \SI{4}{\per\centi\meter}$) for the $E^{'}$ ($A^{'}_{1}$) mode as shown in Fig.~\ref{fig1}d.

Utilizing Stopping and Range of Ions in Matter (SRIM) calculations (see Supporting Information), we are able to compare the sputter yield in our work and the one reported in Ref.~16 to link the He$^{+}$ ion dose to an interdefect distance $L_{D}$ as shown by the upper axis in Fig.~\ref{fig1}c and Fig.~\ref{fig1}d. The same distance metric is used as a reference in the optical and valleytronic characterization measurements discussed in the following.

%
%
%###################### Figure 3 ################################################
\begin{figure}[!ht]
\scalebox{\figurescale}{\includegraphics[width=0.9\linewidth]{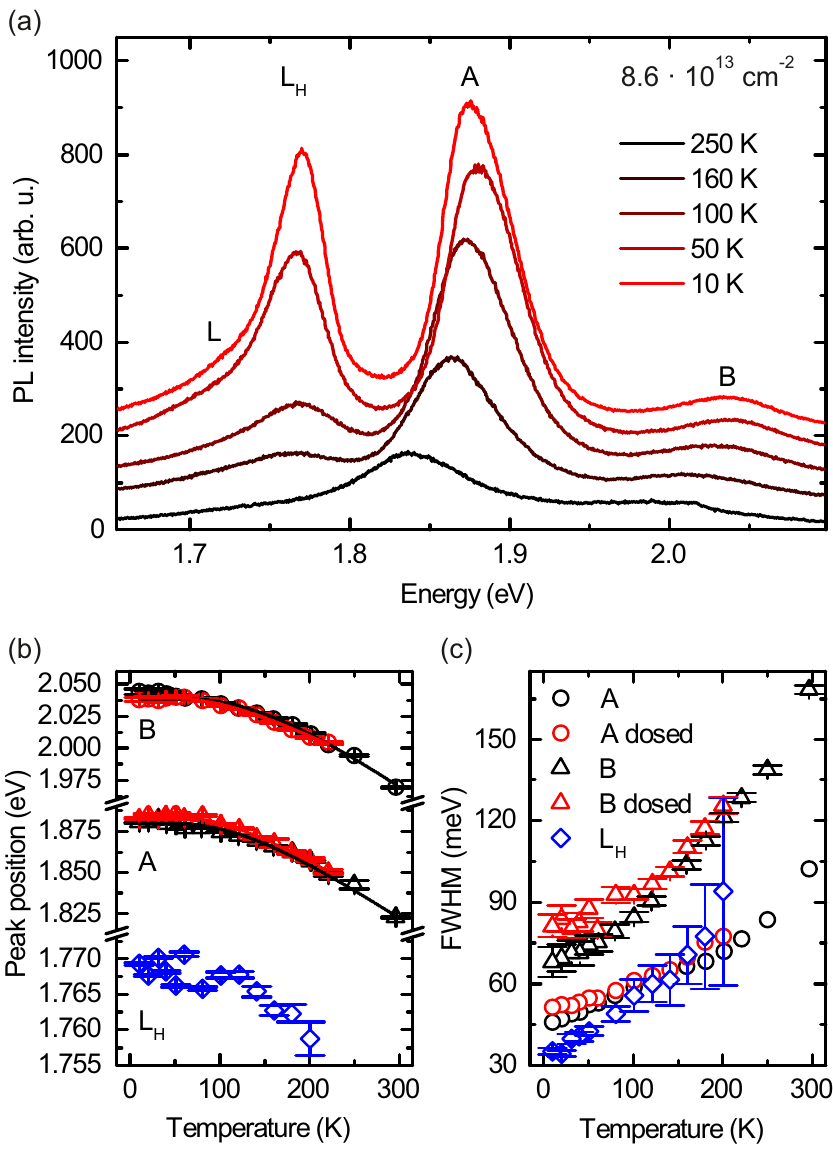}}
\renewcommand{\figurename}{Fig.}
\caption{\label{fig3}
\textbf{Temperature dependence of HIM structured MoS$_{2}$.}
(a) Temperature dependent PL for $\sigma = 8.6 \cdot 10^{13}$ $\SI{}{\per\centi\meter\squared}$ featuring the $A$- and $B$-peaks as well as the $L$- and $L_{H}$-peak.
(b) Temperature dependent peak position for the pristine $A$- (black triangles) and $B$-peaks (black circles) and for the ion-exposed $A$- (red triangles) and $B$-peaks (red circles) and the $L_{H}$ (blue diamonds) peak.
(c) Temperature dependent linewidth for the pristine $A$- (black triangles) and $B$-peaks (black circles) and for the ion-exposed $A$- (red triangles) and $B$-peaks (red circles) and the $L_{H}$ (blue diamonds) peak.
}
\end{figure}

\textbf{Helium-ion modified luminescence in MoS$_{2}$.} To obtain information about the He$^{+}$ induced modification of the MoS$_{2}$ luminescence properties, we perform low-temperature ($T = \SI{10}{\kelvin}$) confocal micro-photoluminescence ($\mu$-PL) spectroscopy exciting at $E_{L} = \SI{2.33}{\electronvolt}$. $\mu$-PL spectra for an increasing irradiation dose are presented in Fig.~\ref{fig2}a. The pristine monolayer spectrum exhibits well-known direct gap luminescence from the K/K' points of the bandstructure. \cite{Mak.2010,Splendiani.2010} We observe both neutral exciton $X$ and trion $T$ luminescence at energies of $E_{X} \sim \SI{1.947}{\electronvolt}$ and $E_{T} \sim \SI{1.916}{\electronvolt}$ due to the low excitation power densities used. \cite{Cadiz.2016,Wierzbowski.2017} Moreover, we find a broad emission at $E_{L} \sim \SI{1.81}{\electronvolt}$, which is referred to the $L$-peak and generally interpreted as defect-related luminescence in the literature. \cite{Splendiani.2010,Mak.2010,Korn.2011} Notably, we detect slight variations of all peak positions of $\Delta E_{peak} \sim \SI{15}{\milli\electronvolt}$ most likely originating from spatial strain variations \cite{Gomez.2013} which are on the order of $\SI{0.1}{\percent}$ in good agreement with the Raman data discussed above.

For an increased $\sigma$, a new luminescence feature emerges, labelled $L_{H}$ in Fig.~\ref{fig2}a that overlaps energetically with the $L$-peak before the overall luminescence is reduced. The $L_{H}$-peak has a very similar lineshape as $X$ and $T$ and arises $\Delta E \sim \SI{180}{\milli\electronvolt}$ ($\Delta E \sim \SI{150}{\milli\electronvolt}$) redshifted with respect to $X$ ($T$). We use multi-peak fits to quantitatively analyze the dose dependent evolution of all peaks in the spectra as shown in Fig.~\ref{fig2}a. In Fig.~\ref{fig2}b, we plot the sum of the peak areas $I_{A} = I_{X} + I_{T}$ (red) and the He$^{+}$ induced $L_{H}$-peak intensity (blue) as a function of $\sigma$. For $\sigma \sim <  10^{14}$ $\SI{}{\per\centi\meter\squared}$, we observe no significant changes in $I_{A}$ with respect to the pristine monolayer. In contrast, $I_{A}$ quenches for higher $\sigma$. This transition occurs for a dose one order of magnitude lower than the characteristic change in the Raman spectra (compare Fig.~\ref{fig1}). Moreover, a constant spectral weight between $X$ and $T$ is observed for $\sigma$ < $10^{14} \SI{}{\per\centi\meter\squared}$ supporting the idea that a possible n- or p-type doping has no significant effect, in good agreement with Raman data. Intriguingly, the intensity of the $L_{H}$-peak rapidly increases with a maximum at $\sigma \sim 10^{14} \SI{}{\per\centi\meter\squared}$ having a higher peak area as compared to $I_{A}$ for pristine MoS$_{2}$. Similar as for $I_{A}$, we observe a strong reduction of the $L_{H}$ intensity with increasing $\sigma$ comparable to the reduction of $I_{A}$ in PL. We note that the so-called $L$-peak known from literature on pristine samples only shows minor changes upon ion-exposure (see Supporting Information).

%
%
%###################### Figure 4 ################################################
\begin{figure}[!ht]
\scalebox{\figurescale}{\includegraphics[width=1\linewidth]{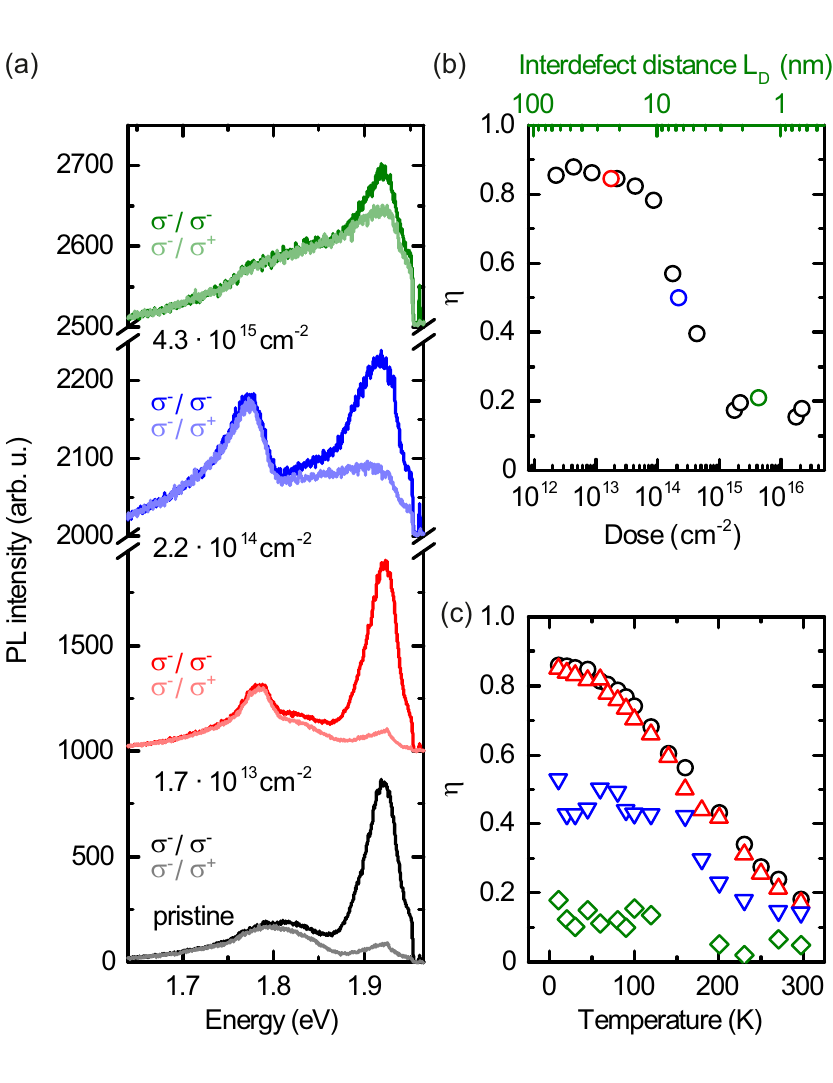}}
\renewcommand{\figurename}{Fig.}
\caption{\label{fig4}
\textbf{Robust valley polarization in HIM structured MoS$_{2}$.}
(a) PL spectra of monolayer MoS$_{2}$ excited by $\sigma^{-}$ and detected with $\sigma^{-}$ (dark) and $\sigma^{+}$ (bright) for pristine and ion-exposed MoS$_2$ with doses of $1.7 \cdot 10^{13}$ $\SI{}{\per\centi\meter\squared}$, $2.2 \cdot 10^{14}$ $\SI{}{\per\centi\meter\squared}$ and $4.3 \cdot 10^{15}$ $\SI{}{\per\centi\meter\squared}$.
(b) Degree of circular polarization $\eta$ of the $A$-peak as a function of $\sigma$.
(c) Temperature dependent degree of circular polarization of pristine and ion-exposed MoS$_2$ for the doses shown in Fig.~\ref{fig4}a.
}
\end{figure}

\newpage

\textbf{Temperature dependent PL spectroscopy.} Temperature dependent PL measurements provide additional insights into the origin and behaviour of the disorder-related peak $L_{H}$. Typical PL spectra for selected temperatures are shown in Fig.~\ref{fig3}a for a sample treated at $\sigma = 8.6 \cdot 10^{13}$ $\SI{}{\per\centi\meter\squared}$ where $L_{H}$-peak luminescence is prominent. Prior to these measurements a higher excitation power density is used for illumination to guarantee that laser induced photodoping effects are negligible for all temperature measurements.\cite{Cadiz.2016,Wierzbowski.2017} The experimental parameters result in a single peak for the $A$-transition that is predominantly composed of trion luminescence. Multi-peak fits reveal the temperature dependence of each feature in the spectrum. Fig.~\ref{fig3}b shows the position of the pristine (black circles) and ion-exposed (red) $A$-peak, pristine (black) and ion-exposed (red) $B$-peak and $L_{H}$-peak (blue diamonds). The temperature dependent peak positions reveal that $L_{H}$ shifts in a very similar manner to $A$ and $B$. Since the $L_{H}$-peak is hardly visible above $\SI{180}{\kelvin}$, fitting is not feasible. Notably, we find no difference between the pristine and ion-exposed peak positions for $A$ and $B$. The temperature dependent linewidth (Fig.~\ref{fig3}c) reveals a strong decrease for decreasing temperatures. In general, MoS$_{2}$ supported on SiO$_{2}$ reveals a significant inhomogeneous linewidth due to spatial inhomogeneities of the substrate. \cite{Ajayi.2017,Wierzbowski.2017} At $\SI{10}{\kelvin}$, the linewidth of the $L_{H}$-peak ($\SI{32}{\milli\electronvolt}$) is sharper as compared to the $A$- ($\SI{45}{\milli\electronvolt}$) and $B$-peak ($\SI{70}{\milli\electronvolt}$) but most probably broadened by a similar mechanism. \\

%
%
%###################### Figure 5 ################################################
\begin{figure}[!ht]
\scalebox{\figurescale}{\includegraphics[width=1\linewidth]{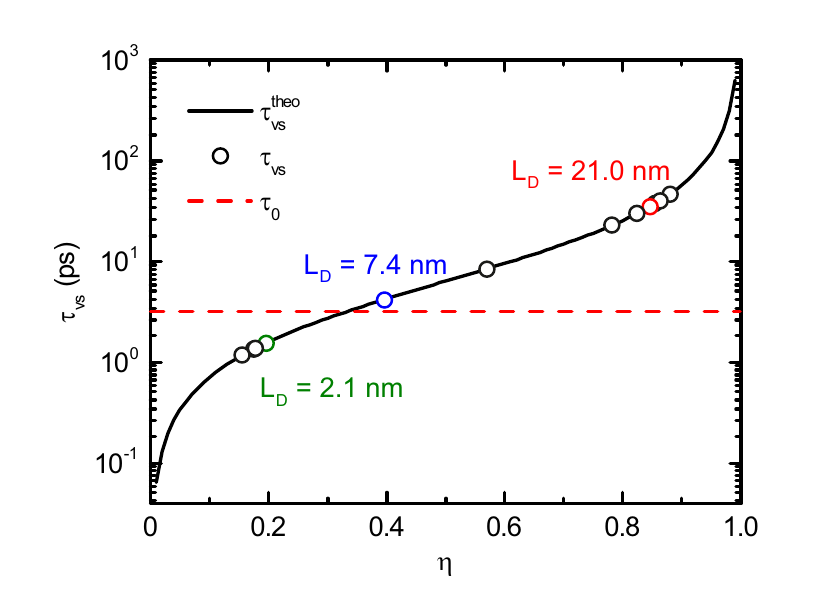}}
\renewcommand{\figurename}{Fig.}
\caption{\label{fig5}
\textbf{Valley spin lifetime in defective MoS$_{2}$.}
Steady state valley spin lifetime as a function of measured valley polarization obtained from Fig.~\ref{fig4}b. The dashed vertical red line indicates the radiative recombination time $\tau_{0} = \SI{4}{\pico\second}$ of the $A$-peak in MoS$_{2}$. \cite{Lagarde.2014}
}
\end{figure}

\textbf{Robust valley polarization.} To obtain information how defects in MoS$_{2}$ affect the valley relaxation rates, we study the degree of circular polarization $\eta = \frac{I_{\sigma_{-}} - I_{\sigma_{+}}}{I_{\sigma_{-}} + I_{\sigma_{+}}}$ for increasing disorder by measuring polarization resolved low-temperature $\mu$-PL. Here, I$_{\sigma^{-}}$ and I$_{\sigma^{+}}$ denote spectra recorded with $\sigma^{-}$ excitation and $\sigma^{-}$ / $\sigma^{+}$ detection as shown in Fig.~\ref{fig4}a for pristine (black) monolayer and ion-exposed material for doses of $1.4 \cdot 10^{13}$ $\SI{}{\per\centi\meter\squared}$ (red), $2.7 \cdot 10^{14}$ $\SI{}{\per\centi\meter\squared}$ (blue) and $5.4 \cdot 10^{15}$ $\SI{}{\per\centi\meter\squared}$ (green). The spectra feature prominent emission from the $A$-peak, $L$-peak and ion-induced $L_{H}$-peak. While the $A$-peak is strongly valley polarized, both, $L$ and $L_{H}$ peak show no circular dichroism for the used excitation at $E = \SI{1.96}{\electronvolt}$ within the given experimental noise.

The corresponding $\eta$ of the $A$-peak as a function of dose is shown in Fig.~\ref{fig4}b.  For pristine MoS$_2$ it exhibits high valley dichroism of $\eta \sim 0.9$, directly reflecting the naturally broken inversion symmetry of the monolayer and low depolarization by intervalley scattering in MoS$_2$ for quasi-resonant excitation of the neutral exciton. For low He$^{+}$ doses $\sigma$ < $10^{14}$ $\SI{}{\per\centi\meter\squared}$ (Fig.~\ref{fig4}a), we observe only a slight drop from $\eta =  0.85$ to $\eta = 0.79$.

The temperature dependent valley polarization of pristine and ion-exposed MoS$_{2}$ with $\sigma = 1.7 \cdot 10^{13}$ $\SI{}{\per\centi\meter\squared}$ (red triangles) show no significant difference with a residual valley polarization of $\eta = 0.2$ at room temperature. Even ion-exposed MoS$_{2}$ with high disorder for $\sigma = 2.2 \cdot 10^{14}$ $\SI{}{\per\centi\meter\squared}$ (blue triangles) reveals a high valley polarization of $\eta \sim 0.5$ at $\SI{10}{\kelvin}$ while it drops to $\eta = \sim 0.15$ for $\sigma = 4.3 \cdot 10^{15}$ $\SI{}{\per\centi\meter\squared}$ (green diamonds).

Based on the spin relaxation for continuous wave excitation \cite{Dyakonov.2008}, we directly translate $\eta$ into a valley spin relaxation time $\tau_{vs}$ using the steady state kinetic Bloch equation $\eta = \frac{1}{1 + 2 \tau_{0} / \tau{vs}}$ where $\tau_{0} = \SI{4}{\pico\second}$ is the radiative recombination time of the $A$-peak. \cite{Lagarde.2014} The valley spin relaxation time as a function of $\eta$ is shown in Fig.~\ref{fig5}. A change in disorder clearly manifests in a strong reduction in $\tau_{vs}$ from $\tau_{vs}(\eta = 0.86) \sim \SI{60}{\pico\second}$ to $\tau_{vs}(\eta = 0.16) \sim \SI{1}{\pico\second}$ to an increased intervalley scattering. This reduction abruptly starts as the interdefect distance approaches the typical radial extension of the free excitons ($\sim\SI{2}{\nano\meter}$).

\section{Discussion}

The experimentally observed opposite shifting of $A^{'}_{1}$ and $E^{'}$ for $\sigma$ > $10^{14}$ $\SI{}{\per\centi\meter\squared}$ in Raman measurements (Fig.~\ref{fig1}c) upon He-ion irradiation closely resembles the activation of Raman scattering in the vicinity of the $\Gamma$ point. We therefore attribute the observed dependence in Fig.~\ref{fig1} to the spread of the phonons in k-space which corresponds to a phonon confinement in real space which has recently been applied to Mn$^{+}$\cite{Mignuzzi.2015} and Ar$^{+}$\cite{Shi.2016} bombarded monolayer MoS$_{2}$. Our observation is in good agreement with the DFT calculated phonon dispersion relation (the full phonon dispersion relation is shown in the Supporting Information) in Fig.~\ref{fig1}e. This is tentatively attributed to increased disorder caused by larger defect clusters; most probably promoted by a strongly increasing number of Mo related defects.\cite{Bae.2017} This interpretation is supported by the sudden decrease of the Raman peak areas for $\sigma$ > $10^{15}$ $\SI{}{\per\centi\meter\squared}$ (see Supporting Information) and changes in morphology as observed by atomic force microscopy (see Supporting Information). The decrease suggests that the phonons spread in momentum space resulting in an increased sampling of phonon frequencies around the $\Gamma$ point. In contrast, we do not observe significant shifts for $\sigma$ < $10^{14}$ $\SI{}{\per\centi\meter\squared}$ where predominantly smaller defects such as monosulphur vacancies $V_{1S}$ with a much lower density are introduced. \cite{Fox.2015}

Moreover, defect-activated Raman modes from the M point at $\SI{361.8}{\per\centi\meter}$, $\SI{374.7}{\per\centi\meter}$ and $\SI{411.2}{\per\centi\meter}$ emerge (green lines in Fig.~\ref{fig1}b) for a typical fit at a dose of $8.61 \cdot 10^{14}$ $\SI{}{\per\centi\meter\squared}$.\cite{Mignuzzi.2015} This observation is interpreted to reflect the breakdown of the Raman selection rule (q $\neq$ 0) combined with a high PDOS at the M point as shown in Fig.~\ref{fig1}e, as it is typically observed in defective TMDCs (see Supporting Information for 1L WSe$_2$). \cite{Mignuzzi.2015,Shi.2016,Stanford.2016,Iberi.2016}  \\

It is well known that the exposure of TMDCs to different ionic species, electrons, annealing or plasma treatments generate chalcogen vacancies in the crystal lattice, since defect formation energies for displacing chalcogen atoms are significantly lower as compared to displacing transition metals. \cite{Komsa.2012} Thus, an increase in $\sigma$ to $10^{14} \SI{}{\per\centi\meter\squared}$ is expected to result in a steadily increasing density of monosulphur vacancies $V_{1S}$ in the crystal.\cite{Bae.2017} A strong reduction of PL intensity is observed for $\sigma$ > $10^{14} \SI{}{\per\centi\meter\squared}$ which as a consequence very likely results in enhanced Auger recombination through the introduction of high disorder in the crystal. This interpretation is in good agreement with large mode shifts in this dose range ($\sigma$ > $10^{14} \SI{}{\per\centi\meter\squared}$) in the Raman data. This is most likely due to an interdefect distance $L_{D} < \SI{10}{\nano\meter}$ that approaches the size of the free excitons in MoS$_{2}$. Generally for low doses ($\sigma$ < $10^{14} \SI{}{\per\centi\meter\squared}$) where mainly V$_{1S}$ are created, there exists an increasing probability for various atoms and molecules that naturally occur in the atmosphere to be chemisorbed\cite{Li.2016} to such mono-sulphur vacancies, generated by the He$^{+}$ ions. To gain further insight into the origin of the modified luminescence, we perform comparative ab-initio DFT calculations to derive the band structure of 1L MoS$_{2}$ with mono- and di-sulphur vacancies with various chemisorbed atoms (C, N, Si, O) and molecules (H$_{2}$O, NH$_{3}$, O$_{2}$, N$_{2}$, CO, OH, NO) to compare with our experiment (see Supporting Information). Due to the fact that we neglect excitonic effects for the calculations, we expect small deviations due to varying binding energies are expected. However, since we only compare relative and not absolute energies, we expect that these variations mostly compensate and that our experimental findings are qualitatively well reproduced. For the calculations considered, we obtain altered direct gap energies ranging from $\Delta E = E_{gap}^{defect} - E_{gap}^{pristine} = \SI{50}{\milli\electronvolt}$ to $\SI{-210}{\milli\electronvolt}$. Comparing with experimental data indicates that defect morphologies which are closest to $\Delta E = \SI{180}{\milli\electronvolt}$ are $O_{2}$ ($\Delta E_{O_{2}} = \SI{210}{\milli\electronvolt}$), Si ($\Delta E_{Si} = \SI{130}{\milli\electronvolt}$), OH ($\Delta E_{OH} = \SI{100}{\milli\electronvolt}$) and NO ($\Delta E_{NO} = \SI{100}{\milli\electronvolt}$) with reasonable defect binding energies $> \SI{2}{\electronvolt}$. The chemisorption of these substitutes at $V_{1S}$ is very likely at ambient atmosphere (in particular for O$_{2}$).

The slight variations for low $\sigma$ < $10^{14}$ $\SI{}{\per\centi\meter\squared}$ in the spectral weight are due to the high intrinsic n-doping of $\sim 10^{12} \SI{}{\per\centi\meter\squared}$ as given by the splitting of the $A^{'}_{1}$ and $E^{'}$ Raman modes.\cite{Miller.2015} This observation is in contrast for previous findings of physisorbed atoms and molecules where a strong shift in spectral weight between neutral and charged exciton has been observed. \cite{Tongay.2013,Tongay.2013b,Nan.2014,Chow.2015} We can estimate the expected change in n-type doping of the MoS$_{2}$ since the total effective charge transfer of O$_{2}$ (Fig.~\ref{fig2}e) by determining the number of defects $N$ at $L_{D} = \SI{10}{\nano\meter}$ ($\sigma =  10^{14} \SI{}{\per\centi\meter\squared}$) where the $L_{H}$-peak area is highest by $\Delta n = N \cdot \Delta Q = \frac{1}{L_{D}^{2}} \cdot 0.8 = 8 \cdot 10^{11} \SI{}{\per\centi\meter\squared}$, which is similar to the intrinsic n-doping for the MoS$_{2}$ sample investigated. Thus, no strong variation in relative spectral weight between $X$ and $T$ are expected to occur for doses $\sigma < 10^{14} \SI{}{\per\centi\meter\squared}$, fully consistent with our data. This argument applies equally to all computed defect morphologies (see Supporting Information). Notably, a control sample with hBN below the ion-bombarded MoS$_{2}$ shows an equivalent additional $L_{H}$-peak, which in our understanding, strongly indicateds that substitutional atoms unlikely stem from the SiO$_{2}$ substrate. All data presented in the main manuscript are from monolayer MoS$_{2}$ on SiO$_{2}$/Si substrate. For all possible substitutes stemming from the gaseous environment, our calculus suggests that O$_{2}$ has the lowest $E_{gap}$ (Fig.~\ref{fig2}c) with still a sufficient binding energy of $\SI{2.81}{\electronvolt}$ for the optical transitions (Fig.~\ref{fig2}d). We find that our calculated binding energies for molecules are in good agreement with recent work. \cite{Li.2016} Future structural studies like TEM, STM and XPS in combination with advanced theoretical calculations, including excitonic effects, are required to unambigiuously confirm our interpretation. However, such structural investigations are beyond the scope of this manuscript. \\

The degree of circular polarization of the emission is closely connected to the symmetry properties of TMDCs. It directly reflects the intravalley and intervalley relaxation rates for monolayer TMDCs with parabolic band disperions at the K/K' points. \cite{Mak.2012,Cao.2012,Zeng.2012} Contributions from short-range and long-range Coulomb interactions, are both found to play a dominant role in the depolarization of the spins. Moreover, the presence of defects is expected to modify the intrinsic scattering rates due to the changes in the crystal potential and due to Coulombic long-range interaction of charged defects (for n-type MoS$_{2}$) \cite{Kaasbjerk.2016}. So far, no thorough study has been conducted on how crystal defects in MoS$_{2}$ affect the intrinsic valley polarization of the $A$-peak. \\

As pointed out before, for such low doses we expect predominantly $V_{1S}$ which correlate with the additional disorder-related luminescence $L_{H}$ observed. Due to the $C_{3v}$ symmetry of $V_{1S}$ almost no contribution to intervalley scattering is expected due to symmetry protection, which is also valid for adatoms.\cite{Kaasbjerk.2016} Indeed for $\sigma$ < $10^{14}$ $\SI{}{\per\centi\meter\squared}$ where disorder-related luminescence attributed due to $V_{1S}$ is observed, a rather constant $\eta$ is experimentally observed supporting this idea. However, for $\sigma$ > $10^{14}$ $\SI{}{\per\centi\meter\squared}$ where more complex defects with different symmetries are expected to emerge and the probability for $V_{Mo}$ creation increases, a significant drop in $\eta$ is found. The interdefect distance becomes comparable to the size of the free excitons, fostering intervalley scattering through enhanced interaction of phonons and excitons at the K and K' valleys. Furthermore, strong intervalley scattering through short-range interaction which can be mediated by $V_{Mo}$ or more complex defect structures and simply by destruction of the crystal occurs. \cite{Kaasbjerk.2016} \\

\section{Conclusion}

Our findings demonstrate that He$^{+}$ exposure can be utilized for nanoscale functionalization and structuring of MoS$_{2}$. In particular, ion doses are found to be low enough such that luminescence properties and valley polarization of the host crystal stay mostly unaltered giving the opportunity to combine high resolution milling \cite{Fox.2015} and HIM lithography \cite{Winston.2009,Sidorkin.2009} for writing nanoscale valleytronic circuits in monolayer MoS$_{2}$, and more generally in TMDCs. Our results demonstrate how the vibrational, optical and spin valley properties of atomically thin MoS$_{2}$ can be modified by the exposure with a focused beam of helium ions. Robust valley polarization is observed for the free excitons in the nanostructured MoS$_{2}$ up to room temperature even for high irradiation doses. This finding is indicative for mono-sulphur vacancies, which are predominantly induced and do not contribute to intervalley scattering due to their symmetry properties. Future steps could comprise controlled generation of particular defects to deterministically generate single photon emitters by proper defect engineering.

\section{Methods}
\subsection{Sample structure}
We employ the viscoelastic transfer method\cite{CastellanosGomez.2014} to transfer large area ($A \sim \SI{1000}{\micro\meter\squared}$) MoS$_2$ monolayer crystals onto $\SI{285}{\nano\meter}$ SiO$_2$ substrates.

\subsection{Helium ion microscopy}
We use MoS$_2$ monolayers transfered onto SiO$_2$/Si substrates for He$^{+}$ irradiation. A beam current $I = \SI{0.86}{\pico\ampere}$ and a beam energy of $\SI{30}{\kilo\volt}$ is used. Areas with a spatial extent of $\SI{4}{\micro\meter}$ x $\SI{4}{\micro\meter}$ are exposed with a beam spacing of $\SI{5}{\nano\meter}$. The dwell time is adjusted such that doses varying from $\sigma \sim 10^{12} \SI{}{\per\centi\meter\squared}$ - $10^{16} \SI{}{\per\centi\meter\squared}$ are obtained.

\subsection{Raman measurements}
Raman measurements are performed at room temperature with the samples held in vacuum. For excitation, a linearly polarized doubled Nd:YAG cw laser with an energy of $\SI{2.33}{\electronvolt}$ is used. An excitation power density of $\SI{113}{\kilo\watt\per\centi\meter\squared}$ is used for excitation throughout all Raman measurements. No detection polarizer is used to obtain maximum signal from the sample. The laser spot has a diameter of $d_{spot} \sim \SI{1.2}{\micro\meter}$.

\subsection{Fitting of Raman spectra}
For fitting Raman spectra in Fig.~\ref{fig1}b we use pseudo-Voigt profiles $V(x)$ to account for each peak
$V(x) = \nu \cdot L(x) + (1 - \nu) \cdot G(x)$ with $L(x) = \frac{1}{\pi} \cdot \frac{\frac{1}{2} \Gamma_{L} A_{L}}{(x_{L}-x_{0})^{2} + (\frac{1}{2}\Gamma_{L})^{2}}$ as Lorentzian and $G(x) = \frac{2 \sqrt{ln2}}{\Gamma_{G} \sqrt{\pi}} \cdot A \cdot \exp{(-\frac{4 \cdot ln2(x_{G}-x_{0})^{2}}{\Gamma^{2}_{G}})}$ as Gaussian part of the function with the corresponding linewidths $\Gamma_{L}$/$\Gamma_{G}$, peak positions $x_{L}$/$x_{G}$ and peak areas $A_{L}$/$A_{G}$. For fitting we center both Lorentzian and Gaussian peak positions by setting $x_{L} = x_{G}$ and fix the linewidth $\Gamma_{L}$ as obtained from fitting a pristine MoS$_{2}$ spectrum.

\subsection{PL measurements}
For low-temperature confocal $\mu$-PL measurements we keep the sample under vacuum in a He-flow cryostat. For helical light resolved measurements, we use a HeNe laser with an excitation energy of $\SI{1.96}{\electronvolt}$ quasi-resonantly with the neutral exciton at $\SI{10}{\kelvin}$. An excitation power density of $\sim \SI{12}{\kilo\watt\per\centi\meter\squared}$ is used. The laser spot has a diameter of $d_{spot} \sim \SI{1.2}{\micro\meter}$. For non-helical light excitation, an excitation energy of $\SI{2.33}{\electronvolt}$ is used with a power density of $\sim \SI{10}{\kilo\watt\per\centi\meter\squared}$ unless otherwise stated in the main text. The laser spot has a diameter of $d_{spot} \sim \SI{1.4}{\micro\meter}$.

\subsection{DFT calculations}
All electronic and geometric calculations are performed using DFT as implemented in the Crystal09 software.\cite{Crystal.09} We employ the all-electron Gaussian-type bases: H5-11G*, C6-21G*, N6-21G*, O8-411, Si88-31G*, S86-311G*, while Mo atoms are treated with the HAYWSC-311(d31)G basis with effective core potential set, together with the PBE gradient corrected density functional.\cite{Perdew.1996}
London-dispersion interactions are accounted for using the approach proposed by Grimme (DFT-D3).\cite{Grimme.2006} The systems are fully optimized, including lattice vectors and atomic positions. The lattice parameters of a perfect MoS$_2$ monolayer are $a=b= \SI{3.178}{\angstrom}$, while for the mono-sulfur-vacancy system, the lattice shrinks to $a = \SI{3.163}{\angstrom}, b = \SI{3.164}{\angstrom}$. The defective systems are calculated using $5\times5$ supercells using 2D periodic boundary conditions, to minimise the interactions between the neighbouring defect sites, and with $2\times2$ k-point mesh. \\
The phonon dispersion curves are obtained for the perfect MoS$_{2}$ monolayer from the VASP code\cite{Kresse.1996, Blchl.1994, Kresse.1999} using projector-augmented waves (PAW) with the energy cutoff of $\SI{600}{\electronvolt}$.
We use the PBE functional.\cite{Perdew.1996} The monolayer is calculated with a $\SI{20}{\angstrom}$ vacuum, to insure negligible interactions with the neighbouring cells in the 3D periodic boundary conditions.
The dispersion corrections are treated at the D3 level. $6\times6\times1$ k-point mesh is used for all the VASP calculations.

%
%##############################################################################
%               Acknowledgements & Contributions
%##############################################################################
%
\section{Acknowledgements}
Supported by Deutsche Forschungsgemeinschaft (DFG) through the TUM International Graduate School of Science and Engineering (IGSSE). We gratefully acknowledge financial support of the German Excellence Initiative via the Nanosystems Initiative Munich and the PhD program ExQM of the Elite Network of Bavaria. A.K. acknowledges ZIH Dresden for computational support and Deutsche Forschungsgemeinschaft (DFG) for Grant GRK 2247/1 (QM3). A.K. thanks Dr. Yandong Ma for his help on VASP.

\section{Author contributions}
J.K., J.J.F., F.K., U.W., A.W.H. and M.K. conceived and designed the experiments, F.S. prepared the samples, J.K. and M.A. performed He-ion exposure of samples, J.K., A.N. and  J.W. performed the optical measurements, J.K. and A.N. analysed the data, A.K. performed and analyzed the DFT calculations, J.K. wrote the manuscript with input from all coauthors.

\section{Abbreviations}
HIM, helium ion microscope; PDOS, phonon density of states; DFT, density functional theory; TMDC, transition metal dichalcogenide; 2D, two-dimensional; PL, photoluminescence;

%###############################################################################
%								Additional information
%###############################################################################
%
\section{Additional information}

\subsection{Supplementary Information} accompanies this paper
\subsection{Competing financial interests} The authors declare no competing financial interests.\\

\textbf{Data availability.} The authors declare that the data supporting the findings of this study are available within the article and its Supplementary Information files. And all data are available from the authors upon request.

%
%###############################################################################
%								BIBLIOGRAPHY
%##############################################################################
%
%\FloatBarrier
%\bibliographystyle{achemso}
\bibliographystyle{apsrev4-1}
\bibliography{Main_Bibliography}% Produces the bibliography via BibTeX.

\end{document}